\documentclass[twocolumn,showpacs,preprintnumbers,amsmath,amssymb]{revtex4}

\usepackage{graphicx}
\usepackage{dcolumn}
\usepackage{bm}
\usepackage[usenames]{color}

\begin{document}


\title{Magnetically induced ferroelectricity in Cu$_{2}$MnSnS$_{4}$ and Cu$_{2}$MnSnSe$_{4}$}

\author{Tetsuya Fukushima}
\email{tetsuya.fukushima@aquila.infn.it}
\author{Kunihiko Yamauchi}
\author{Silvia Picozzi}
\affiliation{%
Consiglio Nazionale delle Ricerche - Superconducting and Innovative Devices (CNR-SPIN), 67100 L'Aquila, Italy\\
}%

\date{\today}
\begin{abstract}

We investigate magnetically-induced ferroelectricity in Cu$_{2}$MnSnS$_{4}$ 
by means of Landau theory of phase transitions and of {\it ab initio} density functional theory. 
As expected from the Landau approach, {\it ab initio} calculations show that a non-zero ferroelectric polarization ${\mbox{\boldmath $P$}}$ along the $y$ direction (of the order of a tenth of $\mu {\rm C}/{\rm cm}^2$) is induced by the peculiar antiferromagnetic configuration of Mn spins occurring in Cu$_{2}$MnSnS$_{4}$. 
The comparison between ${\mbox{\boldmath $P$}}$, calculated either via density-functional-theory or according to Landau approach, 
clearly shows that ferroelectricity is mainly driven by Heisenberg-exchange terms and only to a minor extent by relativistic terms. At variance with previous examples of collinear antiferromagnets with magnetically-induced ferroelectricity (such as AFM-E HoMnO$_3$),  the ionic displacements occurring upon magnetic ordering are very small, so that the exchange-striction mechanism ({\em i.e.} displacement of ions so as to minimize the magnetic coupling energy) is not effective here. Rather, the microscopic mechanism at the basis of polarization has mostly an electronic origin. In this framework, we propose the small magnetic moment at Cu sites induced by neighboring Mn magnetic moments to play a relevant role in inducing ${\mbox{\boldmath $P$}}$. Finally, we investigate the effect of the anion by comparing Cu$_{2}$MnSnSe$_{4}$ 
and Cu$_{2}$MnSnS$_{4}$: Se-4$p$ states, more delocalized compared to S-3$p$ states, are able to better mediate the Mn-Mn interaction, in turn leading to a higher ferroelectric polarization in the Se-based compound.
\end{abstract}

\pacs{75.50.E, 77.80.-e, 75.85.+t}
\maketitle

\section{\label{sec:intro}Introduction\protect\\}
Multiferroic materials have attracted both scientific and industrial interests, due to their profound physics and novel functionalities, respectively.\cite{Cheong} Among them, many studies have been recently performed for ``improper'' ferroelectric materials,\cite{Kimura,Hur,Picozzi,Sergienko,Yamauchi} where ferroelectricity is driven by non-centrosymmetric spin, charge, or orbital ordering. In this case, a stronger magnetoelectric coupling is expected, compared to the conventional proper ferroelectrics. So far, mainly manganites (such as TbMnO$_{3}$, TbMn$_{2}$O$_{5}$, and HoMnO$_{3}$\cite{Kimura,Hur,Picozzi,Sergienko}) and other perovskite oxides\cite{Gianluca1,Sanjeev} have been carefully investigated by means of theoretical and experimental approaches, in order to better understand  cross-coupling effects between spin/orbital orderings and dielectric properties.

The search for novel multiferroics other than oxides has already started: recently N\'{e}nert {\it et al.}\cite{Nenert} have suggested the ternary copper chalcogenide Cu$_{2}$MnSnS$_{4}$ (with nominal valences as ${\rm Cu}^{1+}$, ${\rm Mn}^{2+}$, ${\rm Sn}^{4+}$, and ${\rm S}^{2-}$) as a candidate of ``improper'' ferroelectricity, using a symmetry-based  analysis and the Landau phenomenological theory, with the ferroelectricity induced by a peculiar antiferromagnetic (AFM) Mn spin configuration. 
Experimentally, neutron diffraction measurements have clarified that such AFM configuration occurs under $T_{N}$=8.8 K, with Mn-spins slightly deviating from the $c$ axis by an angle between 6$^{\circ}$ and 16$^{\circ}$ degree. Moreover,
the magnetization curve has suggested the spin quantization axis to be close to the $c$ direction.\cite{Fries}
Additionally, the small dependence of susceptibility upon the direction of magnetic field implied a rather small magnetic anisotropy.  

Similar to the famous stannite, Cu$_{2}$FeSnS$_{4}$,\cite{Caneschi} Cu$_{2}$MnSnS$_{4}$ crystallizes in the tetragonal sphalerite superstructure, which shows the $I\bar{4}2m$ (No.121) space group with symmetry operations \{$E$, $C_{2x}$, $C_{2y}$, $C_{2z}$, $\sigma_{d\bar{x}}$, $\sigma_{d\bar{y}}$, $S_{4}$, $S^{-1}_{4}$\}.\cite{symmetry} The crystal structure is similar to zincblende, often seen in III-V and II-VI type semiconductors, where each cation is tetrahedrally coordinated to  anions.  In this respect, multiferroic effects in this class of materials might prove useful in the context of semiconductor-based spintronics.
Although the sphalerite crystal structure itself lacks inversion symmetry, this does not automatically imply that the crystal is polar: due to the mirror and $\pi$-rotation symmetries, the ferroelectric polarization ${\mbox{\boldmath $P$}}$ is forbidden along any direction. 
On the other hand, N\'{e}nert {\it et al.} have claimed that the experimentally observed AFM spin order breaks some symmetries, so that ${\mbox{\boldmath $P$}}$ arises along the $y$ direction ({\em i.e.} the crystallographic $b$ direction) with possibly large magnetoelectric coupling.\cite{Nenert}
The
Landau phenomenological theory, based on the symmetry analysis of both the crystal structure and the magnetic order, is a powerful tool to investigate  magnetically-induced ferroelectricity.\cite{Sergienko,Toledano}
However, in order to quantify the value of ${\mbox{\boldmath $P$}}$ and to clarify the relation between magnetism and ferroelectricity via the electronic structure, a combination of Landau theory and {\it ab initio} density functional theory (DFT) calculations is desirable.\cite{Picozzi}
\begin{figure}[t]
\begin{center}
\includegraphics[width=9.5cm,clip]{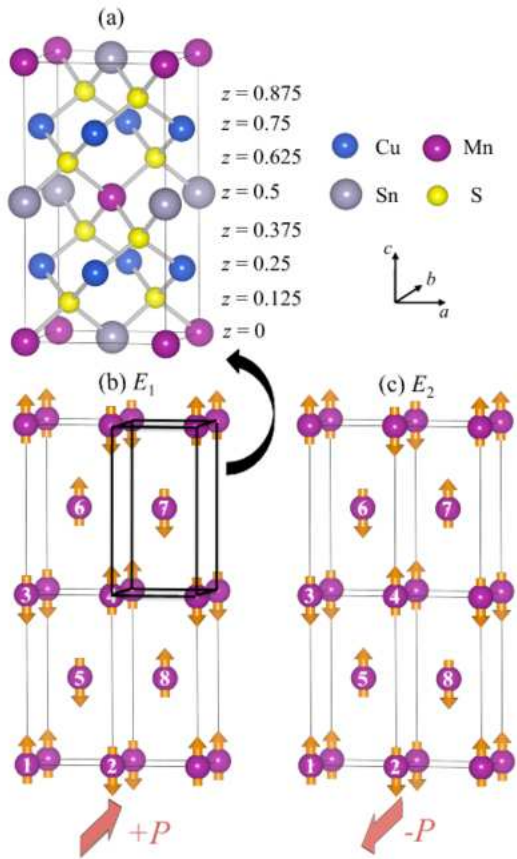}
\caption{(a) Nuclear crystal structure of Cu$_{2}$MnSnS$_{4}$. (b) and (c) indicate AFM unit cells with the propagation vector $\mbox{\boldmath $k$}=\left(\frac{1}{2},0,\frac{1}{2}\right)$ corresponding to ferroelectric domains with positive ${\mbox{\boldmath $P$}}$ and negative -${\mbox{\boldmath $P$}}$, respectively. For simplicity, only Mn atoms are shown in (b) and (c). Orange arrows at Mn sites indicate the spin directions. The magnetic unit cell is doubled along the $a$ and $c$ directions, compared to the nuclear unit cell. Labels of Mn spins are shown in each sphere.}
\label{crystal}
\end{center}
\end{figure}
In this paper, we investigate magnetically-induced ferroelectricity in Cu$_{2}$MnSnS$_{4}$ by using both Landau theory and DFT in the aim of carefully understanding the microscopic mechanism at the basis of the multiferroic behaviour.  
To check the effect of the anions on ${\mbox{\boldmath $P$}}$, we perform similar simulations by substituting S with Se, $i.e.$ for Cu$_{2}$MnSnSe$_{4}$.  
First, we will discuss the direction and the value of ${\mbox{\boldmath $P$}}$ obtained by DFT, confirming the predictions based on Landau theory; afterwards,
we will focus on the ferroelectric switching, in terms of different Mn-spin configurations along the adiabatic path between  positive and negative ferroelectric states.
Finally, the microscopic mechanism leading to ferroelectricity
via spatially-polar Cu and Mn charge densities caused by Heisenberg-exchange is discussed. 
\section{Methodology and structural details}
DFT calculations are performed for Cu$_{2}$MnSnS$_{4}$ and Cu$_{2}$MnSnSe$_{4}$ by using the ``Vienna Ab initio Simulation Package'' (VASP), based on Projector Augmented Wave (PAW) pseudopotentials.\cite{Kresse} The Perdew-Becke-Erzenhof (PBE) approach to the generalized gradient approximation (GGA) is employed for the exchange-correlation potential.\cite{Perdew} The plane wave energy cut-off is 400 eV for each atom. During structural optimization, a threshold on the atomic forces is set as 0.01 eV/${\rm \AA}$. Internal atomic coordinates are optimized starting from experimental data of Cu$_{2}$MnSnS$_{4}$\cite{Fries} and Cu$_{2}$MnSnSe$_{4}$.\cite{Sachanyuk} The $2{\times}4{\times}1$ Monkhost-Pack $k$-point grid in the Brillouin zone is used.\cite{Monkhorst} Cu-$3d4s$, Mn-$3d4s$, Sn-$5s5p$, S-$3s3p$, and Se-$4s4p$ electrons are treated as valence states. The GGA+$U$ calculations within the Dudarev's approach\cite{Dudarev} are performed by applying a Hubbard-like potential only for Mn-$d$ states. An on-site Coulomb parameter $U$ = 4.0 eV and an exchange parameter $J$ = 0.89 eV are used. 

Lattice constants are fixed to the experimental data: $a=b=5.514$ \AA, $c=10.789$ ${\rm \AA}$  for Cu$_{2}$MnSnS$_{4}$\cite{Fries} and $a=b=5.736$ \AA, $c=11.401$ ${\rm \AA}$ for Cu$_{2}$MnSnSe$_{4}$\cite{Sachanyuk} in the tetragonal $I\bar{4}2m$ structure. 
In order to impose the experimentally observed AFM spin configuration in Cu$_{2}$MnSnS$_{4}$\cite{Fries}  with the propagation vector ${\mbox{\boldmath $k$}}=(\frac{1}{2}, 0, \frac{1}{2})$, we build the  magnetic super-cell with 8 formula/units ({\em i.e.} 64 atoms/cell, see Fig.\ref{crystal}(b) and (c))
(note that, to our knowledge, the experimental magnetic configuration for Cu$_{2}$MnSnSe$_{4}$ has not been reported yet).
Since the observed AFM configuration breaks the symmetries (except for $C_{2y}$), ferroelectric polarization is allowed to be magnetically-induced along the $y$ direction.
To evaluate ${\mbox{\boldmath $P$}}$, we also perform calculations for the ferromagnetic (FM) spin configuration, which is considered as a reference non-polar state:  the ferroelectric polarization is then calculated as $\Delta{\mbox{\boldmath $P$}}={\mbox{\boldmath $P$}}_{\rm AFM}-{\mbox{\boldmath $P$}}_{\rm FM}$. 
To connect paraelectric (PE) and ferroelectric (FE) states, calculations for noncollinear spin configurations (see below) are also performed according to Ref.\cite{Hobbs}. The Berry phase approach developed by King-Smith and Vanderbilt\cite{Vanderbilt}  is employed to calculate the electric polarization ${\mbox{\boldmath $P$}}$, where we integrate over eight $k$-point strings parallel to ${\mbox{\boldmath $P$}}$.

\section{Magnetically induced ferroelectric polarization}
\subsection{Landau theory}
In the Landau potential terms, the
ferroelectric polarization ${\mbox{\boldmath $P$}}$ appears as coupled with magnetic order parameters, in terms
which are invariant under symmetry operations.\cite{Mostovoy,Harris} 
Nen\'{e}rt {\it et al.}\cite{Nenert} have built up the free energy equation by considering two Mn spins (${\mbox{\boldmath $S$}}_{1}$ with position (0,0,0) and ${\mbox{\boldmath $S$}}_{2}$ with position ($\frac{1}{2}$,$\frac{1}{2}$,$\frac{1}{2}$) in the nuclear unit cell)
 and set up two magnetic order parameters, {\it e.g}., the FM order parameter, ${\mbox{\boldmath $M$}}={\mbox{\boldmath $S$}}_{1}+{\mbox{\boldmath $S$}}_{2}$, and the AFM order parameter, ${\mbox{\boldmath $L$}}={\mbox{\boldmath $S$}}_{1}-{\mbox{\boldmath $S$}}_{2}$. 
From the free energy, they have showed that only the $P_{y}$ component is coupled with both $L_{x}$ and $L_{z}$, 
and also with $M_{y}$.  
Thereby, they have concluded that a spontaneous polarization arises along the $y$ direction upon the AFM configuration. 
Here we extend their model to fit into our AFM supercell with eight Mn spins, 
so that we can discuss the adiabatic switching path between the positive FE ($+{\mbox{\boldmath $P$}}$) and negative FE ($-{\mbox{\boldmath $P$}}$) states by changing the direction of Mn spins and the consequent electric polarization.  (The same procedure was applied for HoMnO$_{3}$ with E-type AFM configuration.\cite{Picozzi, Sergienko})
With these eight Mn spins (labelled as ${\mbox{\boldmath $S$}}_{1},\cdots,{\mbox{\boldmath $S$}}_{8}$, as in Fig.\ref{crystal}), two AFM order parameters are constructed as
\begin{eqnarray}
{\mbox{\boldmath $E$}}_{1}&=&{\mbox{\boldmath $S$}}_{1}-{\mbox{\boldmath $S$}}_{2}-{\mbox{\boldmath $S$}}_{3}+{\mbox{\boldmath $S$}}_{4}-{\mbox{\boldmath $S$}}_{5}+{\mbox{\boldmath $S$}}_{6}-{\mbox{\boldmath $S$}}_{7}+{\mbox{\boldmath $S$}}_{8}, \nonumber \\ {\mbox{\boldmath $E$}}_{2}&=&{\mbox{\boldmath $S$}}_{1}-{\mbox{\boldmath $S$}}_{2}-{\mbox{\boldmath $S$}}_{3}+{\mbox{\boldmath $S$}}_{4}+{\mbox{\boldmath $S$}}_{5}-{\mbox{\boldmath $S$}}_{6}+{\mbox{\boldmath $S$}}_{7}-{\mbox{\boldmath $S$}}_{8},
\end{eqnarray}
corresponding to the positive and negative ferroelectric domain phases with $+{\mbox{\boldmath $P$}}$ and $-{\mbox{\boldmath $P$}}$, respectively. 
In order to switch between the two phases (as seen in the following section), four body-centered Mn spins (${\mbox{\boldmath $S$}}_{5}$, ${\mbox{\boldmath $S$}}_{6}$, ${\mbox{\boldmath $S$}}_{7}$, ${\mbox{\boldmath $S$}}_{8}$) must be flipped. 
Taking into account the irreducible corepresentation of the little group of $I\bar{4}2m$ with the propagation vector  ${\mbox{\boldmath $k$}}=(\frac{1}{2}, 0, \frac{1}{2})$, one can write the possible magnetoelectric coupling terms in the second degree of magnetic order parameters and the dielectric term in the Landau potential as: 
\begin{eqnarray} \label{LF2}
F_{\rm ME}&=&c_{0}{\mbox{\boldmath $E$}}^{2}_{1}P_{y}+c'_{0}{\mbox{\boldmath $E$}}^{2}_{2}P_{y}+c^{xx}_{1}{E^{x}_{1}}^{2}P_{y}+c^{xx}_{2}{E^{x}_{2}}^{2}P_{y} \nonumber \\ &\ & +c^{zz}_{1}{E^{z}_{1}}^{2}P_{y}+c^{zz}_{2}{E^{z}_{2}}^{2}P_{y}+c^{xx}_{12}E^{x}_{1}E^{x}_{2}P_{y} \nonumber \\ &\ &+c^{xz}_{11}E^{x}_{1}E^{z}_{1}P_{y}+c^{xz}_{12}E^{x}_{1}E^{z}_{2}P_{y}+c^{xz}_{21}E^{x}_{2}E^{z}_{1}P_{y} \nonumber \\ [0.5mm] &\ &+c^{xz}_{22}E^{x}_{2}E^{z}_{2}P_{y}+c^{zz}_{12}E^{z}_{1}E^{z}_{2}P_{y} \nonumber \\ &\ & +\frac{1}{2}(\chi^{-1}_{x}P_{x}^{2}+\chi^{-1}_{y}P_{y}^{2}+\chi^{-1}_{z}P_{z}^{2}),
\end{eqnarray}
where the $c$ coefficients  denote phenomenological parameters and $\chi_{x}$, $\chi_{y}$ and $\chi_{z}$ are the components of the dielectric susceptibility. Since Mn magnetic moments have been experimentally observed in the $xz$ plane, the terms containing $E_{1y}$ and $E_{2y}$ are omitted in the above equation. The first and second terms denote exchange terms, the last term is the dielectric energy and the remaining terms are relativistic terms. The minimization of Eq.(\ref{LF2}) gives a finite polarization along $y$: 
\begin{eqnarray}\label{POL2}
P_{y}&=&-\chi_{y}(c_{0}{\mbox{\boldmath $E$}}^{2}_{1}+c'_{0}{\mbox{\boldmath $E$}}^{2}_{2}+c^{xx}_{1}{E^{x}_{1}}^{2}+c^{xx}_{2}{E^{x}_{2}}^{2} \nonumber \\ &\ & +c^{zz}_{1}{E^{z}_{1}}^{2}+c^{zz}_{2}{E^{z}_{2}}^{2}+c^{xx}_{12}E^{x}_{1}E^{x}_{2} \nonumber \\ &\ &+c^{xz}_{11}E^{x}_{1}E^{z}_{1}+c^{xz}_{12}E^{x}_{1}E^{z}_{2}+c^{xz}_{21}E^{x}_{2}E^{z}_{1} \nonumber \\ [0.5mm] &\ &+c^{xz}_{22}E^{x}_{2}E^{z}_{2}+c^{zz}_{12}E^{z}_{1}E^{z}_{2}),
\end{eqnarray}
whereas non-zero $P_{x}$ and $P_{z}$ are not obtained. 
This result confirms N\'{e}nert's predictions: a spontaneous $P_{y}$ is evidently induced by the AFM coupling. However, at this point we cannot conclude whether the Heisenberg-exchange- or the relativistic-interaction terms is dominant in  inducing $P_{y}$.

\subsection{DFT results}
%
\begin{figure}
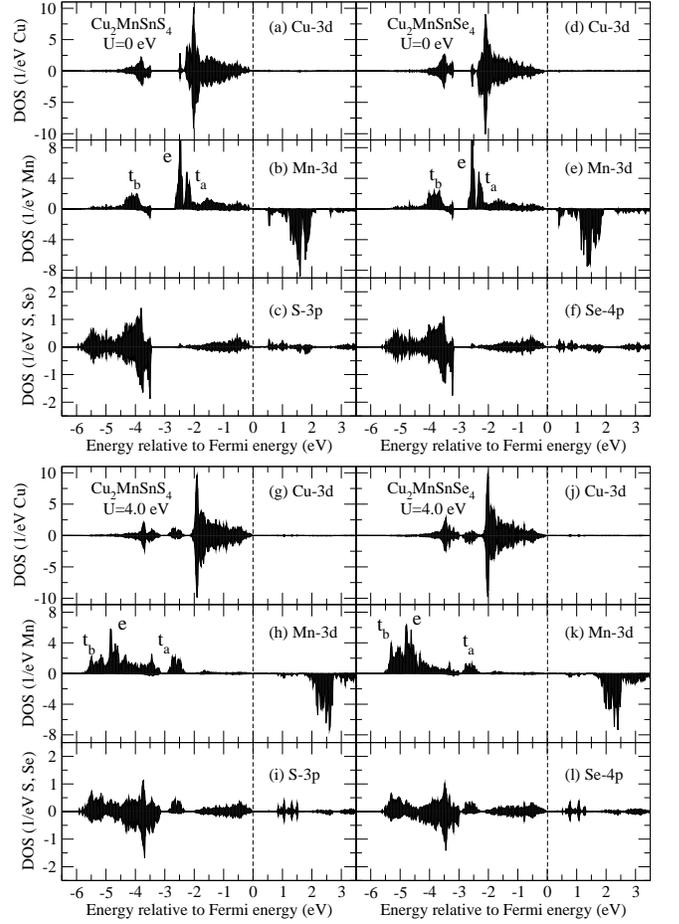

\begin{center}
\begin{tabular}{c}
{\includegraphics[width=8.5cm,clip]{DOS_u_0.eps}} \\
{\includegraphics[width=8.5cm,clip]{DOS_u_4.eps}} \\
\end{tabular}
\caption{The upper panel shows Cu-3$d$ (a), Mn-3$d$ (b) and S-3$p$ (c) partial DOSs of AFM Cu$_2$MnSnS$_4$ and Cu-3$d$ (d), Mn-3$d$ (e) and Se-4$p$ (f) partial DOSs of AFM Cu$_2$MnSnSe$_4$ (at $U$ = 0 eV). The lower panel shows the case of $U$ = 4.0 eV. The zero of the energy scale marks the Fermi energy.
}
\label{DOS}
\end{center}
\end{figure}
Figure \ref{DOS} shows the partial density of states (DOS) projected onto (a) Cu-$3d$, (b) Mn-$3d$ and (c) S-$3p$ states in AFM Cu$_{2}$MnSnS$_{4}$ and onto (d) Cu-$3d$, (e) Mn-$3d$ and (f) Se-$4p$ states in AFM Cu$_{2}$MnSnSe$_{4}$. To discuss the electronic structure, we focus on Cu$_{2}$MnSnS$_{4}$. 
The valence bands just below the Fermi level mainly come from Cu-3$d^{10}$ states with small contribution from Mn-3$d^{5}$ states, while the bottom of conduction bands is mainly from Mn-3$d$ states. We recall that, since Mn atoms are tetrahedrally surrounded by S atoms, the five-fold degenerate 3$d$ states split into the three-fold $t_{2}$ states and two-fold $e$ states due to the crystal electric field (CEF). Additionally, the three-fold degenerated $t_{2}$ states form bonding $t_{b}$ and anti-bonding $t_{a}$ states. 
The majority spin states are fully occupied whereas the minority spin states are fully unoccupied, corresponding to a high-spin $d^5$ configuration (with 4.4 $\mu_{\rm B}$ magnetic moment inside the atomic sphere with 1.3 \AA\: radius).  
Cu-3$d$ states are weakly spin polarized with a small induced magnetic moment (equal to 0.015 $\mu_{\rm B}$ inside the atomic sphere with 1.3 \AA\: radius).
We note, in fact, that the distance between Mn atomic sites is rather large, 
so that localized Mn-spins interact via Cu and S sites. 
Fig.\ref{DOS} (a) and (b) show that there is a small hybridization between Cu-$3d$ and Mn-$3d$ states, so that  Cu-ions are slightly spin-polarized. As we'll show later, the small Cu spins caused by the Cu-Mn hybridization play an important role for ferroelectricity in these materials.

%
\begin{table} \begin{center} \begin{tabular}{ccccccccc} 
\hline
&$U$ (eV)&SOC& $P_{x}$& $P_{y}$  & $P_{z}$\\ \hline 
Cu$_{2}$MnSnS$_{4}$&0 & on, $E//x$ & 0 &  0.209& 0 \\
					 && on, $E$//z & 0 &  0.208& 0 \\
                                & & off & 0 &  0.209& 0 \\
                                & 4.0& on, $E$//x &0 & 0.054 & 0 \\
					 && on, $E$//z &0 & 0.054 & 0 \\
					 && off&0 & 0.055 & 0 \\
Cu$_{2}$MnSnSe$_{4}$& 0& on, $E//x$ & 0 &  0.339& 0 \\
					& & on, $E$//z & 0 &0.335  & 0 \\
                                & & off & 0 & 0.339 & 0 \\
                                               & 4.0& on, $E$//x &0 & 0.092 & 0 \\
					 && on, $E$//z &0 & 0.092 & 0 \\
					 && off&0 & 0.095 & 0 \\
 \hline
\end{tabular}
\caption{DFT-calculated ${\mbox{\boldmath $P$}}$  (${\mu}{\rm C}/{\rm cm}^{2}$) of AFM Cu$_{2}$MnSnS$_{4}$ and Cu$_{2}$MnSnSe$_{4}$ for $U$ = 0 and 4.0 eV, upon switching on and off SOC. When including SOC, all Mn spins are aligned along the $x $ or $z$ direction. }
\label{polarization} \end{center} \end{table}
Table \ref{polarization} shows the DFT calculated ${\mbox{\boldmath $P$}}$ in the optimized structure, imposing the AFM configuration. 
Consistent with Landau theory results, only $P_{y}$ has a finite value.
Both compounds show a rather small ${\mbox{\boldmath $P$}}$ compared to other materials with magnetically-induced ferroelectric polarization (comparing with ${\mbox{\boldmath $P$}}$ equal to few $\mu {\rm C}/{\rm cm}^{2}$ in E-type AFM HoMnO$_{3}$\cite{Picozzi}, one or two orders of magnitude smaller); this is likely due to the weaker ``indirect" interaction (via Mn-S(Se)-Cu-S(Se)-Mn bond).
 
The reason why Cu$_{2}$MnSnSe$_{4}$ shows a larger ${\mbox{\boldmath $P$}}$ than Cu$_{2}$MnSnS$_{4}$ has to be traced back to the fact that Mn-$3d$ states can hybridize more with  Se-$4p$ states than with (more localized) S-$3p$ states, in turn mediating a stronger interaction between Mn-$3d$ and Cu-$3d$ states. In the same respect, we note that the difference between $U$ = 0 and 4.0 eV as far as ${\mbox{\boldmath $P$}}$ is concerned can be equivalently understood in terms of Mn-3$d$ and Cu-3$d$ hybridization: the on-site Coulomb interaction pushes Mn-3$d$ states  deeper in energy (clearly shown in Fig.\ref{DOS}) and decreases the Mn-3$d$ and Cu-3$d$ hybridization. Therefore, ${\mbox{\boldmath $P$}}$ values at $U$ = 4.0 eV are smaller compared to the cases of $U$ = 0 eV.

The important finding here is that  the contribution to ${\mbox{\boldmath $P$}}$ due to spin-orbit coupling (SOC) is very small, as shown by the comparison of polarization values switching on or off the relativistic interactions ({\it cf.} Table \ref{polarization}). 
In this respect, we also note that the total energy decreases - by 0.04 meV/Mn - when the spin direction is changed  from the $a$ to $c$ axis, in agreement with experiments.

We recall that our Landau theory analysis didn't clarify which magnetic term ($i.e.$ Heisenberg exchange or relativistic) mainly contributes to ${\mbox{\boldmath $P$}}$; on the other hand,
DFT results unambiguously show the relativistic term to give only a minor contribution.
Therefore we can omit the terms depending on either $E^{x}$ or $E^{z}$ in Eq. (\ref{POL2}), but keep the ${\mbox{\boldmath $E$}}^{2}$ term so that the equation is simplified as:
\begin{eqnarray}\label{POL3}
P_{y}&=&-\chi_{y}(c_{0}{\mbox{\boldmath $E$}}^{2}_{1}+c'_{0}{\mbox{\boldmath $E$}}^{2}_{2}) \nonumber \\ &=& (-\chi_{y}c_{0}-\chi_{y}c'_{0})\sum_{i{\neq}j}{\mbox{\boldmath $S$}}_{i}{\cdot}{\mbox{\boldmath $S$}}_{j}.
\end{eqnarray}
%
\begin{figure}
\begin{center}
\begin{tabular}{c}
{\includegraphics[width=8.5cm,clip]{new_path2.eps}} \vspace{0.5cm} \\
{\includegraphics[width=8cm,clip]{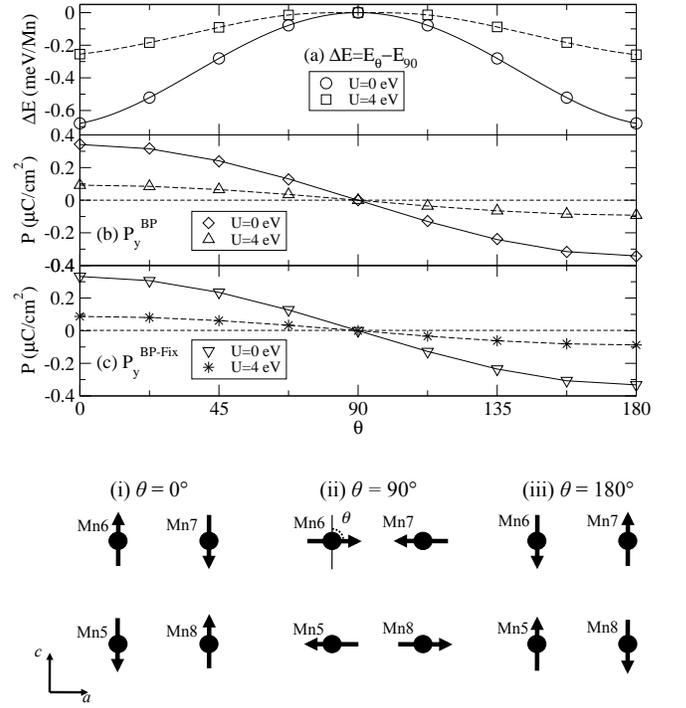}} \\
\end{tabular}
\caption{(a) Total energy differences ${\Delta}E=E-E_{\theta=90^{\circ}}$ as a function of the relative angle $\theta$ at $U$ = 0 eV and $U$ = 4.0 eV. (b) and (c) show $P_{y}$ calculated by the Berry phase method for the optimized atomic structure and for   the atomic structure fixed to the case of $\theta=90^{\circ}$, respectively. The lower panel shows the spin directions of Mn5, Mn6, Mn7 and Mn8 atoms (labels consistent with Fig.\ref{crystal}).}
\label{NONCOL}
\end{center}
\end{figure}
Here, $P_{y}$  depends only on the inner products of Mn spin vectors, therefore is described by the relative spin angle. ${\mbox{\boldmath $S$}}_{i}{\cdot}{\mbox{\boldmath $S$}}_{i}$ terms in the above equation are neglected because these terms disapper when calculating the polarization difference between FE and PE states.
Progressive rotation of (${\mbox{\boldmath $S$}}_{5}$, ${\mbox{\boldmath $S$}}_{6}$, ${\mbox{\boldmath $S$}}_{7}$, ${\mbox{\boldmath $S$}}_{8}$) Mn spins can switch the ferroelectric polarization between $+{\mbox{\boldmath $P$}}$ and $-{\mbox{\boldmath $P$}}$ continuously.
Assuming the  non-collinear (NC) spin configuration as ${\mbox{\boldmath $S$}}_{1}=-{\mbox{\boldmath $S$}}_{2}=-{\mbox{\boldmath $S$}}_{3}={\mbox{\boldmath $S$}}_{4}=(0,0,1)$ and $-{\mbox{\boldmath $S$}}_{5}={\mbox{\boldmath $S$}}_{6}=-{\mbox{\boldmath $S$}}_{7}={\mbox{\boldmath $S$}}_{8}=(\sin\theta,0,\cos\theta)$, Eq.(\ref{LF2}) becomes
\begin{equation}\label{POL4}
P_{y}=-\chi_{y}c_{0}(32\cos\theta+32)-\chi_{y}c'_{0}(-32\cos\theta+32).
\end{equation}
%
Indeed, the  $P_{y}\propto \cos{\theta}$ trend is obtained when we perform DFT calculations on the NC configuration by  varying $\theta$ between 0 and 180$^{\circ}$ with fully optimized atomic coordinates for each spin configuration.
Figure \ref{NONCOL} shows the calculated $P^{\rm BP}_{y}$ and the total energy  difference ${\Delta}E(\theta)=E(\theta)-E(\theta=90^{\circ})$ along the adiabatic path as a function of the relative Mn-spin angle $\theta$ at Cu$_{2}$MnSnSe$_{4}$. We also show $P^{\rm BP-Fix}_{y}$, calculated with atomic coordinates fixed to the case of $\theta=90^{\circ}$, so that there is no contribution to the electric polarization from atomic displacements. The similarity between polarization values for Fig.\ref{NONCOL} (b) and (c) - including or not the ionic displacements - highlights stannite as a paradigmatic example of {\em purely electronic} ferroelectricity (see also discussion below). 

As expected from Landau theory and as pointed out above, 
the DFT-calculated $P^{\rm BP}_{y}$ is fitted by a cosine curve, with coefficients $-\chi_{y}c_{0}=\chi_{y}c'_{0}=$0.0053 and 0.0015 at $U$= 0 and 4.0 eV, respectively. 
The function ${\Delta}E$ shows convex and symmetrical behavior between positive and negative FE states. 
With increasing $\theta$ from $\theta=0^{\circ}$ (positive FE state), ${\Delta}E$ increases and reaches maximum value at $\theta=90^{\circ}$ (PE state) with the energy barrier ${\Delta}E(0){\simeq}$ 0.68 and 0.25 meV/Mn at $U$= 0 and 4.0 eV, respectively. This energy barrier ---coming from combination of $J_{ij}{\mbox{\boldmath $S$}}_{i}{\cdot}{\mbox{\boldmath $S$}}_{j}$ term and ${\mbox{\boldmath $P$}}$-related term---
  is much lower than in the HoMnO$_{3}$ case, where ${\Delta}E(0)$ = 8 meV/Mn at $U$ = 0 eV.\cite{Picozzi}
  This is fully consistent with the weak Mn-Mn exchange-coupling constants involved.%
\section{Mechanism of ferroelectricity}
\begin{figure}[t]
\begin{center}
{\includegraphics[width=8cm,clip]{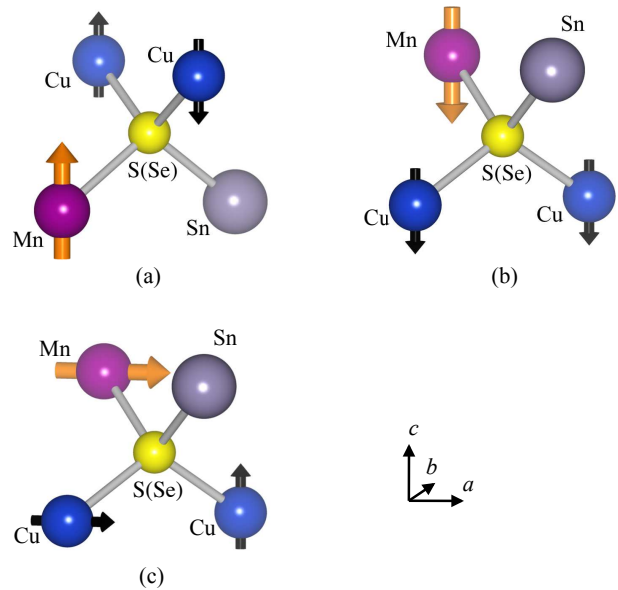}}
\caption{Schematics of S(Se) sites tetrahedrally surrounded by one Mn, two Cu, and one Sn atoms in the $I\bar{4}2m$ symmetry. Sn atoms do not have magnetic moment. The orange (large) and black (small) arrows indicate spins of Mn and Cu atoms, respectively.}
\label{tetra}
\end{center}
\end{figure}
According to the previous section, ferroelectric polarization is induced by Heisenberg-exchange coupling between Mn spins.  We also noted that the Heisenberg exchange between collinear spins is more efficient in driving ferroelectric polarization than any spin-orbit Dzyaloshinskii-Moriya interaction\cite{Mostovoy}: the order of magnitude of ${\mbox{\boldmath $P$}}$ here (tenth or hundredth of $\mu {\rm C}/{\rm cm}^2$) is about one order of magnitude higher than
that experimentally observed and calculated (hundredth or thousandths of $\mu {\rm C}/{\rm cm}^2$) in spiral-based systems such as TbMnO$_3$. 
The origin of ferroelectricity could be explained by analogy with the HoMnO$_{3}$ case, 
where
O$^{2-}$ ions move in order to reduce the magnetic coupling energy between Mn spins.
Such ``exchange-striction" mechanism has been referred to as ``inverse Goodenough-Kanamori'' (iGK) interaction.\cite{Picozzi,Yamauchi2}
However, since here in Cu$_2$MnSnS$_4$ the Mn ions are far apart, the Mn spin interaction is too ``indirect" and it is difficult 
 (and non-intuitive) to guess how S(Se) ions would move to reduce the magnetic coupling energy. Indeed, as shown in Fig.\ref{NONCOL} (b) and (c), $P^{\rm BP}_{y}$ and $P^{\rm BP-Fix}_{y}$ have almost the same value: the ionic displacements are rather small and do not contribute much to ${\mbox{\boldmath $P$}}$. 

\begin{figure}[t]
\begin{center}
{\includegraphics[width=8cm,clip]{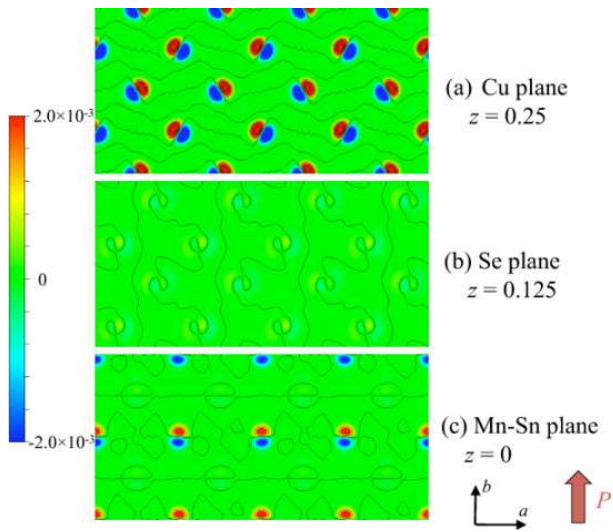}}
\caption{$ab$ plane charge density difference between the FE and PE states in Cu$_{2}$MnSnSe$_{4}$: $\Delta\rho({\mbox{\boldmath $r$}})=\rho_{\rm FE}({\mbox{\boldmath $r$}})-\rho_{\rm PE}({\mbox{\boldmath $r$}})$. (a), (b) and (c) show the differences of charge density distributions at $z$=0, 0.125 and 0.25 planes ({\it cf.} Fig.\ref{crystal}(a)), respectively.}
\label{chg}
\end{center}
\end{figure}

Rather, to explain the  ``electronic" origin of the Heisenberg-driven polarization, we assume small-size spins (actually calculated as 0.015 $\mu_{\rm B}$) on Cu sites, which interact with Mn spins via ``Cu-S(Se)-Mn'' bonds. Consider S(Se) sites: from the chemical point of view, all of them are equivalent, since they are tetrahedrally surrounded by one Mn, two Cu and one non-magnetic Sn atoms in the symmetry $I\bar{4}2m$. However, from the magnetic point of view, due to the small magnetic moments at Cu sites, S(Se) ions can have two different magnetic environments at the FE state ($\theta=0^{\circ}$): the magnetic moments of the surrounding Cu atoms can align either parallel or antiparallel, as shown in Fig.\ref{tetra} (a) and (b). The anions are therefore split into two categories, they become inequivalent upon this peculiar AFM spin ordering, breaking the symmetries and paving the way to ferroelectric polarization. Since Mn spins interact via such magnetically inequivalent S(Se) ions, the charge densities at Mn sites are modulated by Heisenberg-exchange to induce ${\mbox{\boldmath $P$}}$. Similarly, charge densities at Cu sites with small magnetic moments caused by Mn-$3d$ and Cu-$3d$ hybridization can polarize. This is clearly observed in Fig.\ref{chg}, where we show the difference between FE and PE states: Mn and Cu charge densities are clearly non-symmetric along $b$, {\em i.e.} ${\mbox{\boldmath $P$}}$ along $y$ is induced. The results are consistent with the magnetically broken point group 2 at Mn and Cu sites (the point groups at Cu and Mn sites were ``originally" - {\em i.e.} without magnetic ordering -  $-42m$ and $-4$, respectively). The ``original" site symmetry at S(Se) sites is $m$, which is already a polar point group; therefore, a magnetically induced polar charge does not appear. We note that, in the PE state ($\theta=90^{\circ}$), 
some Cu magnetic moments point along the $x$ direction, corresponding to the NC Mn spin configuration. All S(Se) ions have then an equivalent magnetic environment, where half of the Cu magnetic moments align parallel to Mn and the other half perpendicular to it
(Fig.\ref{tetra} (c)); in this configuration, ${\mbox{\boldmath $P$}}$ is therefore not expected along any direction.

\section{Summary}

We have investigated magnetically-induced ferroelectricity in AFM Cu$_{2}$MnSnS$_{4}$ and Cu$_{2}$MnSnSe$_{4}$ by means of Landau theory and DFT calculations. The combination of both methods 
clearly highlights that the driving force of ${\mbox{\boldmath $P$}}$ is not the relativistic term, but the Heisenberg exchange term. Microscopically, small localized Cu magnetic moments induced by Mn magnetic moments play an important role in inducing ${\mbox{\boldmath $P$}}$. The Mn-Mn spin interaction via the magnetically inequivalent S(Se) ions produces a polar modulation of the charge density on Mn and Cu sites, finally inducing a net ${\mbox{\boldmath $P$}}$ with predominantly {\em electronic} contribution and negligible contribution from {\em ionic displacements}. We therefore show that AFM Cu$_{2}$MnSnS$_{4}$ and Cu$_{2}$MnSnSe$_{4}$ are remarkable examples of  novel magnetically-induced ferroelectric materials, with improper multiferroicity branching out into non-oxide and non-octahedral based systems.

\acknowledgments
The research leading to these results has received funding from the European Research Council under the European Community's 7th Framework Programme (FP7/2007-2013) / ERC grant agreement n. 203523. Computational support by CASPUR supercomputing center (Rome) is acknowledged. TF thanks G. N\'{e}nert for helpful discussions and  for his careful reading of the manuscript.
The crystal structures and charge densities in this paper are plotted by using a software VESTA.\cite{VESTA}


\begin{thebibliography}{200}
\bibitem{Cheong}
S.-W. Cheong and M. Mostvoy, Nat. Mater. {\bf 6} (2007) 13.
\bibitem{Kimura}
T. Kimura, T. Goto, H. Shintani, K. Ishizaka, T. Arima, and Y. Tokura, Nature (London) {\bf 426} (2003) 56.
\bibitem{Hur}
N. Hur, S. Park, P. A. Sharma, J. S. Ahn, S. Guha, and S. W. Cheong, Nature (London) {\bf 429} (2004) 392.
\bibitem{Picozzi}
S. Picozzi, K. Yamauchi, B. Sanyal, I. A. Segienko, and E. Dagotto, Phys. Rev. Lett. {\bf 99} (2007) 227201.
\bibitem{Sergienko}
I. A. Sergienko, C. Sen, and E. Dagotto, Phys. Rev. Lett. {\bf 97} (2006) 227204.
\bibitem{Yamauchi}
K. Yamauchi, F. Freimuth, S. Bl\"{u}gel, and S. Picozzi, Phys. Rev. B {\bf 78} (2008) 014403.
\bibitem{Gianluca1}
G. Giovannetti, S. Kumar, D. Khomskii, S. Picozzi, and J. van den Brink, Phys. Rev. Lett. {\bf 103} (2009) 156401.
\bibitem{Sanjeev}
S. Kumar, G. Giovannetti, J. van den Brink, and S. Picozzi, cond-mat/9091439 (unpublished).
\bibitem{Nenert}
G. N\'{e}nert and T. T. M. Palstra, J. Phys.: Condens. Matter {\bf 21} (2009) 176002.
\bibitem{Fries}
T. Fries, Y. Shapira, F. Palacio, M. C. Mor\'{o}n, G. J. McIntyre, R. Kershaw, A. Wold, and E. J. McNiff Jr., Phys. Rev. B {\bf 56} (1997) 5624.
\bibitem{Caneschi}
A. Caneschi, C. Cipriani, F. Di Benedetto, and R. Sessoli, Phys. Chem. Minerals {\bf 31} (2004) 190.
\bibitem{symmetry}
$\sigma_{d\bar{x}}$ and $\sigma_{d\bar{y}}$ denote mirror symmetries with respect to (110) and (1-10) planes, respectively.
\bibitem{Toledano}
P. Tol\'{e}dano, Ferroelectrics {\bf 161} (1994) 257.
\bibitem{Kresse}
G. Kresse and J. Furthmuller, Phys. Rev. B {\bf 54} (1996) 11169.
\bibitem{Perdew}
J. P. Perdew, K. Burke, and M. Ernzerhol, Phys. Rev. Lett. {\bf 77} (1996) 3865.
\bibitem{Sachanyuk}
V. P. Sachanyuk, I. D. Olekseyuk, and O. V. Parasyuk, Phys. Stat. Sol. (a) {\bf 203} (2006) 459.
\bibitem{Monkhorst}
H. J. Monkhorst and J. D. Pack, Phys. Rev. B {\bf 13} (1976) 5188.
\bibitem{Dudarev}
S. L. Dudarev, G. A. Botton, S. Y. Savrasov, C. J. Jumphreys, and A. P. Sutton, Phys. Rev. B {\bf 57} (1998) 1505.
\bibitem{Hobbs}
D. Hobbs, G. Kresse, and J. Hafner, Phys. Rev. B {\bf 62} (2000) 11556.
\bibitem{Vanderbilt}
R. D. King-Smith and D. Vanderbilt, Phys. Rev. B {\bf 47} (1993) 1651.
\bibitem{Mostovoy}
M. Mostovoy, Phys. Rev. Lett. {\bf 96} (2006) 067601.
\bibitem{Harris}
A. B. Harris, T. Yildirim, A. Aharony, and O. Entin-Wohlman, Phys. Rev. B {\bf 73} (2006) 184433.
\bibitem{Yamauchi2}
K. Yamauchi and S. Picozzi, J. Phys.: Condens. Matter {\bf 21} (2009) 064203.
\bibitem{VESTA}
K. Momma and F. Izumi, J. Appl. Crystallogr. {\bf 41} (2008) 653.
\end{thebibliography}

\end{document}